\documentclass[aps,pra,twocolumn,superscriptaddress,nofootinbib]{revtex4}
\usepackage{amsmath,amssymb,bm}
\usepackage[pdftex]{graphicx}
\usepackage{txfonts}
\usepackage{ulem}
\usepackage{color}
\newcommand{\dell}{\Delta'}
\newcommand{\dnu}{\Delta''}
\newcommand{\delltwo}{\Delta^{\!(2)\prime}}
\newcommand{\dnutwo}{\Delta^{\!(2)\prime\prime}}

\begin{document}

\title{Topological classification of the single-wall carbon nanotube}

\author{Rin Okuyama}
\email{rokuyama_at_meiji_dot_ac_dot_jp}
\affiliation{Department of Physics, Meiji University, Kawasaki 214-8571, Japan}
\affiliation{Faculty of Science and Technology,
	Keio University, Yokohama 223-8522, Japan}

\author{Wataru Izumida}
\affiliation{Department of Physics, Tohoku University,
	Sendai 980-8578, Japan}

\author{Mikio Eto}
\affiliation{Faculty of Science and Technology,
	Keio University, Yokohama 223-8522, Japan}

\date{received 16 December 2018; published 11 March 2019}

\begin{abstract}
The single-wall carbon nanotube (SWNT) can be a
one-dimensional topological insulator, which is characterized by
a $\mathbb{Z}$-topological invariant, winding number.
Using the analytical expression for the winding number,
we classify the topology for all possible chiralities of SWNTs
in the absence and presence of a magnetic field,
which belongs to the topological categories of BDI and AIII,
respectively.
We find that the majority of SWNTs are nontrivial topological insulators
in the absence of a magnetic field. In addition,
the topological phase transition takes place when the band gap is closed
by applying a magnetic field along the tube axis,
in all the SWNTs except armchair nanotubes.
The winding number determines the number of edge states localized at
the tube ends by the bulk-edge correspondence,
the proof of which is given for SWNTs in general.
This enables the identification of the topology in experiments.
\end{abstract}

\maketitle

\section{Introduction}

The single-wall carbon nanotube (SWNT) is a quasi-one-dimensional (1D)
material made by rolling up a graphene sheet, which possesses
two Dirac cones at $K$ and $K'$ points.
The circumference of nanotube is represented by the chiral vector,
$
	\bm{C}_{\rm h} = n \bm{a}_1 + m \bm{a}_2
$,
on the graphene, where $\bm{a}_1$ and $\bm{a}_2$ are
the primitive lattice vectors and
a set of two integers, $(n, m)$, is called chirality
	\cite{Saito1998}.
The SWNT is metallic (semiconducting) for
$
	{\rm mod} (2n + m, 3) = 0
$
$(\neq 0)$,
because some wavevectors discretized in the circumference direction
pass (do not pass) through $K$ or $K'$ points when
they are expressed in the two-dimensional (2D)
Brillouin zone (BZ) of graphene.
Even for metallic SWNTs, a narrow band gap opens due to
the finite curvature in the tube surface
	\cite{Hamada1992, Saito1992, Kane1997}.
The curvature enhances the spin-orbit (SO) interaction
through the mixing between $\pi$ and $\sigma$ orbitals, which also
contributes to the band gap
	\cite{Ando2000}.

Recently, SWNTs have attracted attention from a viewpoint of topology
	\cite{Klinovaja2012, Egger2012, Sau2013, Hsu2015,
	Izumida2016, Lin2016, Okuyama2017, Izumida2017,
	Marganska2018, Zang2018}.
The neutral SWNT can be regarded as a 1D insulator in the
presence of band gap and rotational symmetry (see below).
Due to the sublattice (or chiral) symmetry between $A$ and $B$
lattice sites, 
the topology of a SWNT is characterized by a $\mathbb{Z}$
topological invariant, winding number
	\cite{Wen1989}.
SWNTs can be 1D topological insulators
in both the absence and presence of a magnetic field,
which belong to classes BDI and AIII
in the periodic table in Ref.\
	\cite{Schnyder2009},
respectively.
Izumida {\it et al.} introduced the winding number
for semiconducting SWNTs for the first time
	\cite{Izumida2016}.
They also examined the edge states localized around the tube ends
with energy $E = E_{\rm F} = 0$, the number of which
is related to the winding number by the bulk-edge correspondence.
This enables us to know the winding number via the observation of
local density of states at the tube ends by the
scanning tunneling microscopy as already done for the graphene
	\cite{Kobayashi2005}.
The present authors generalized the theory for metallic SWNTs
	\cite{Okuyama2017}.
The narrow band gap in metallic SWNTs can be closed by applying
a magnetic field of a few Tesla along the tube axis.
This results in the topological phase transition,
where the winding number
changes discontinuously as a function of the magnetic field.
Independently, Lin {\it et al.} examined the topological nature in
a zigzag SWNT ($n > 0$ and $m = 0$)
by using the Su-Schrieffer-Heeger
model and topological invariant called Zak phase
	\cite{Lin2016}.
They theoretically proposed
a possible manipulation of the edge states via
the topological phase transition, although it requires an
unrealistically huge magnetic field
in the case of a semiconducting SWNT.
There also exist theoretical studies on topological phases in
a SWNT proximity coupled to a superconductor
	\cite{Klinovaja2012, Egger2012, Sau2013,
	Hsu2015, Izumida2017, Marganska2018}.

In the present study,
we topologically classify all possible SWNTs.
The winding number is analytically derived
for all possible chiralities.
We also generalize the bulk-edge correspondence to the cases of
both semiconducting and metallic SWNTs in a magnetic field
along the tube axis,
which determines the number of edge states by the winding number.
Our main results are depicted in Fig.\ \ref{fig:class}:
(a) In the absence of a magnetic field,
the majority of SWNTs are topological insulators
with nonzero winding number.
The exceptions are metallic SWNTs of armchair type ($n = m$) and 
semiconducting SWNTs with $n = m + 1$.
(b) In the presence of a magnetic field,
the topological phase transition takes place when the band gap is
closed by applying a magnetic field, for all SWNTs other than
the armchairs. In other words,
the SWNT can be topologically nontrivial
even for $n = m + 1$ when the magnetic field is tuned
appropriately.
Only armchair nanotubes are topologically trivial
regardless of the magnetic field, which is
 due to the mirror symmetry with respect to
a plane including the tube axis
	\cite{Izumida2016}.
Previously, some groups theoretically predicted a change in
the number of edge states in a SWNT as a function of magnetic field
	\cite{Sasaki2005, Sasaki2008, Marganska2011}.
Our theory clearly explains their physical origin in terms of topology.

We noticed a theoretical study by Zang {\it et al.}
	\cite{Zang2018}
during the preparation of this paper.
They utilized a similar technique to ours to analyze
the winding number in SWNTs.
They showed that some SWNTs can have the edge states and that
the topological phase transition takes place by applying a
magnetic field.
Their study, however, was only applicable to semiconducting SWNTs,
and they did not derive analytic expression for the winding number.

This paper is organized as follows.
In Sec.\ \ref{sec:semicond}, we introduce
a 1D lattice model for semiconducting SWNTs in the absence of
a magnetic field,
utilizing the rotational symmetry.
We include a magnetic field along the tube axis in Sec.\ \ref{sec:mag}.
In Sec.\ \ref{sec:analytical}, we analytically evaluate
the winding numbers in the case of semiconducting SWNTs in
both the absence and presence of a magnetic field.
The winding number determines
the number of edge states via the bulk-edge
correspondence, whose proof is given in Appendix \ref{app:EOM}.
In Sec.\ \ref{sec:metal},
we examine the topology in metallic SWNTs with small band gap
induced by the curvature effects.
After the discussion on our theoretical study in Sec.\ \ref{sec:discussion},
our conclusions are given in Sec.\ \ref{sec:conclusions}.

\begin{figure}[t] \begin{center}
\includegraphics{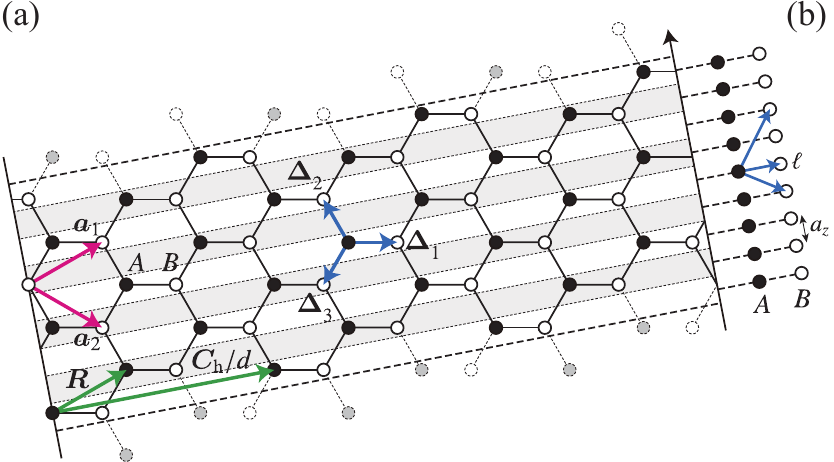}
\caption{
	(Color online) (a) The mapping of the $(n, m)$-SWNT to
	a graphene sheet.
	The chiral vector, $\bm{C}_{\rm h} = n \bm{a}_1 + m\bm{a}_2$,
	indicates the circumference of the tube with
	$\bm{a}_1$ and $\bm{a}_2$ being the primitive lattice vectors of
	graphene.
	The three vectors
	$\bm{\Delta}_j$
	$(j = 1, 2, 3)$
	connect the nearest-neighbor atoms.
	The $d$-fold symmetry around the tube axis (helical symmetry)
	corresponds to the translational symmetry of
	$\bm{C}_{\rm h} / d$ $(\bm{R} = p \bm{a}_1 + q \bm{a}_2)$, where
	$d = \gcd(n, m)$ and integers $p$ and $q$ are given by
	Eq.\ \eqref{eq:pq}.
	This figure shows the case of $(n, m) = (6, 3)$ with
	$d = 3$, $p = 1$, and $q = 0$.
	(b) A 1D lattice model in which $A$ and $B$
	lattice sites are aligned in the axial direction.
}
\label{fig:lattice}
\end{center} \end{figure}

\section{1D lattice model for semiconducting nanotube
	\label{sec:semicond}}

In this section, we derive a 1D lattice model for
semiconducting SWNTs in the absence of a magnetic field.
Neither the Aharonov-Bohm (AB) effect in a magnetic field
nor curvature-induced narrow gap in metallic SWNTs are considered.

Throughout the paper, we consider the $(n, m)$-SWNT,
whose circumference is specified by
chiral vector
$
	\bm{C}_{\rm h} = n \bm{a}_1 + m \bm{a}_2
$
on a graphene sheet,
where $\bm{a}_{1/2} = (\sqrt3/2, \pm 1/2) a$ with the lattice constant
$a = 0.246~{\rm nm}$ [see Fig.\ \ref{fig:lattice}(a)].
Its diameter is given by
$
	d_{\rm t} = |\bm{C}_{\rm h}|/ \pi =
	a \sqrt{n^2 + nm + m^2} / \pi
$.
The chiral angle $\theta$ is defined as
the angle between $\bm{C}_{\rm h}$ and $\bm{a}_1$:
$
	\theta = \tan^{-1} [\sqrt3 m / (2n + m)]
$.
We restrict ourselves to the case of
$
	0 \leq m \leq n
$
without loss of generality, which corresponds to
$
	0 \leq \theta \leq \pi / 6
$
with $\theta = 0$ and $\pi / 6$ for
zigzag ($m = 0$) and armchair ($m = n$)
nanotubes, respectively.

\subsection{Derivation}

We start from the tight-binding model for graphene
	\cite{Saito1998},
which consists of
$A$ and $B$ sublattices,
as depicted by filled and empty circles,
respectively, in Fig.\ \ref{fig:lattice}(a).
This model involves an isotropic hopping integral $\gamma$ between
the nearest-neighbor atoms.
An $A$ atom is connected to three $B$ atoms by vectors
$\bm{\Delta}_j$
$(j = 1, 2, 3)$ in Fig.\ \ref{fig:lattice}(a).
The Hamiltonian reads
\begin{align}
	H = \sum_{\bm{r}_A} \sum_{j = 1}^3 \left(
		\gamma c^{\dagger}_{\bm{r}_A}
		c_{\bm{r}_A + \bm{\Delta}_j}
		+ {\rm H.c.}
	\right),
\end{align}
where
$\bm{r}_\sigma$ is the position of $\sigma = A$ or $B$
atom on the graphene sheet, and
$
	c_{\bm{r}}
$
is the field operator for a $\pi$ electron of atom at
position $\bm{r}$.
$
	\gamma = - 2 \hbar v_{\rm F} / (\sqrt3 a)
$
with
$
	v_{\rm F} = 8.32 \times 10^5 ~ {\rm m/s}
$
being the Fermi velocity in the graphene.
The spin index $s$ is omitted,
which is irrelevant
in Secs.\ \ref{sec:semicond} and \ref{sec:mag}.

We derive a 1D lattice model of
the SWNT along the lines of Ref.\
	\cite{Izumida2016},
where the helical-angular construction
	\cite{White1993, Jishi1993}
is utilized.
$(n, m)$-SWNT has the $d$-fold rotational
symmetry around the tube axis, where
\begin{align}
	d = \gcd(n, m)
		\label{eq:def_d}
\end{align}
is the greatest common divisor of $n$ and $m$.
The rotation by $2\pi/d$ corresponds
to the translation by
$\bm{C}_{\rm h} / d$ on the graphene sheet.
The SWNT
also has the helical symmetry represented by
the translation by
$
	\bm{R} = p \bm{a}_1 + q \bm{a}_2
$
on the graphene sheet,
with integers $p$ and $q$ satisfying
\begin{align}
	mp - nq = d.
		\label{eq:pq}
\end{align}
This means that the SWNT is invariant under the
translation by
$
	a_z = \sqrt3 d a^2 / (2 \pi d_{\rm t})
$
along the tube axis together with the rotation by
$
	\theta_z = 2 \pi [(2n + m) p + (n + 2m) q]
		/ [2(n^2 + nm + m^2)]
$
around it [see Fig.\ \ref{fig:lattice}(a)].\footnote{
	Note that there is an arbitrariness for the choice of
	$p$ and $q$ in Eq.\ \eqref{eq:pq}:
	$\bm{R}$ can be added by integer multiple of
	$\bm{C}_{\rm h} / d$.
	$a_z$ is invariant whereas
	$\theta_z \rightarrow \theta_z \pm 2 \pi / d$
	when $\bm{R} \rightarrow \bm{R} \pm \bm{C}_{\rm h} / d$.
}
Here, $\bm{R}$ and $\bm{C}_{\rm h} / d$ are a new set of
primitive lattice vectors of graphene;
the position of $A$ and $B$ atoms can be expressed as
\begin{align}
	\bm{r}_A &= \ell \bm{R} + \nu (\bm{C}_{\rm h} / d),
		\label{eq:r_A} \\
	\bm{r}_B &= \ell \bm{R} + \nu (\bm{C}_{\rm h} / d)
		+ \bm{\Delta}_1,
		\label{eq:r_B}
\end{align}
on the graphene sheet with site indices $\ell$ and
$
	\nu = 0, 1, 2, \ldots, d - 1
$.
By performing the Fourier transformation for the $\nu$ coordinate,
we obtain the Hamiltonian block diagonalized in
the subspace of orbital angular momentum
$
	\mu = 0, 1, 2, \ldots, d - 1
$
as
$
	H = \sum_{\mu = 0}^{d - 1} H_{\mu}
$,
\begin{align}
	H_{\mu} = \sum_{\ell} \sum_{j = 1}^3 \left( \gamma
		{\rm e}^{{\rm i} 2 \pi \mu \dnu_j / d}
		c^{\, \mu \, \dagger}_{A, \ell}
		c^{\, \mu}_{B, \ell + \dell_j}
		+ {\rm H.c.} \right).
		\label{eq:H0}
\end{align}
This is a 1D lattice model in which $A$ and $B$
lattice sites are aligned in the axial direction
with the lattice constant $a_z$,
as shown in Fig.\ \ref{fig:lattice}(b).
Here, $c^{\, \mu}_{\sigma, \ell}$ is the field operator of
an electron with angular momentum $\mu$ and
at sublattice $\sigma$ of site index $\ell$.
The hopping to the $j$th nearest-neighbor atom
[vector in $\bm{\Delta}_j$ in Fig.\ \ref{fig:lattice}(a)]
gives rise to the hopping to
the sites separated by
$\dell_j$
in Fig.\ \ref{fig:lattice}(b) with phase factor
$\dnu_j$,
where 
\begin{align}
	\bm{\Delta}_j - \bm{\Delta}_1
	= \dell_j \bm{R} + \dnu_j (\bm{C}_{\rm h} / d),
		\label{eq:Delta}
\end{align}
or $\dell_1 = \dnu_1 = 0$,
$\dell_2 = n / d$, $\dnu_2 = -p$,
$\dell_3 = - m/d$, and $\dnu_3 = q$.

\subsection{Bulk properties}

For the bulk system,
the Fourier transformation of $H_{\mu}$ along
$\ell$ direction yields the two-by-two Hamiltonian,
\begin{align}
	H_{\mu} (k) &= \gamma \begin{bmatrix}
		0 & f_\mu (k) \\
		f^*_\mu (k) & 0
	\end{bmatrix},
		\label{eq:H0_k}
\end{align}
in the sublattice space for given wave number $k$.
$k$ runs through the 1D BZ,
$
	-\pi \leq k a_z < \pi
$,
and
\begin{align}
	f_\mu (k) &= \sum_{j = 1}^3 
		{\rm e}^{{\rm i} 2 \pi \mu \dnu_j / d}
		{\rm e}^{{\rm i} k a_z \dell_j}.
		\label{eq:f0}
\end{align}
The dispersion relation for subband
$\mu$ is readily obtained as
\begin{align}
	E_{\mu} (k) &= \pm |\gamma f_\mu (k)|.
		\label{eq:epsilon0}
\end{align}
The system is an insulator for semiconducting SWNTs with
${\rm mod}(2n + m, 3) \neq 0$, that is,
$
	f_\mu (k) \neq 0
$
in the whole BZ.
Then, the positive and negative $E_{\mu} (k)$'s form
the conduction and valence bands, respectively.
On the other hand, for metallic SWNTs with
${\rm mod}(2n + m, 3) = 0$, $f_\mu (k)$ becomes
zero at $\mu_+$ and $k_+$ ($\mu_-$ and $k_-$)
that correspond to the $K$ ($K'$) point on the graphene sheet,
as discussed in Sec.\ \ref{sec:metal}.

\subsection{Winding number and bulk-edge correspondence}

The bulk Hamiltonian in Eq.\ \eqref{eq:H0_k} anticommutes with
$\sigma_z$ in the sublattice space, which is called sublattice
or chiral symmetry.
Thanks to this symmetry as well as the finite band gap,
we can define the winding number
	\cite{Wen1989},
\begin{align}
	w_\mu = \int_{-\pi/a_z}^{\pi/a_z} \frac{{\rm d} k}{2 \pi}
		\frac{\partial}{\partial k} \arg f_\mu (k)
	\equiv \frac{1}{2 \pi} \oint_{\rm BZ} {\rm d} \arg f_\mu (k),
		\label{eq:w}
\end{align}
for subband with angular momentum $\mu$ in semiconducting SWNTs
	\cite{Izumida2016}.
The winding number is the number of times that
$f_\mu (k)$ in Eq.\ \eqref{eq:f0} winds around
the origin on the complex plane when $k$ runs through the 1D BZ.
Note that $w_\mu$ in Eq.\ \eqref{eq:w} is ill-defined for
metallic SWNTs where $f_\mu (k)$ is zero and
therefore $\arg f_\mu (k)$ cannot be defined at $\mu_\tau$ and $k_\tau$
($\tau = \pm 1$).
We will overcome this problem in Sec.\ \ref{sec:metal}.

The bulk-edge correspondence holds between
the winding number and number of edge states, $N_{\rm edge}$,
\begin{align}
	N_{\rm edge}
	= 4 \sum_{\mu = 0}^{d - 1}
		|w_\mu|
		\label{eq:bulk_edge}
\end{align}
in a long but finite SWNT.
The prefactor of 4 is
ascribable to the spin degeneracy and two edges at tube ends.
This relation was analytically shown for semiconducting SWNTs in
the absence of a magnetic field in Ref.\ \cite{Izumida2016}.
We generalize Eq.\ \eqref{eq:bulk_edge}
[and Eq.\ \eqref{eq:bulk_edge_metal}]
for both semiconducting
and metallic SWNTs in
a magnetic field in Appendix \ref{app:EOM}.
Here, we assume that the tube is cut by a broken line in
Fig.\ \ref{fig:lattice}(a),
which results in so-called minimal boundary edges
	\cite{Akhmerov2008}.
The case of the other boundaries is discussed in
Sec.\ \ref{sec:discussion}.

\section{1D lattice model with finite magnetic field
	\label{sec:mag}}

We extend our theory to include a magnetic field $B$
in the axial direction of the SWNT.
We neglect the spin-Zeeman effect throughout the paper,
which is justified unless the band gap is closed
in a huge magnetic field.\footnote{The large Zeeman effect
could overlap the conduction band for one spin and valence
band for the other spin, which makes the system metallic.}
Only the AB effect is taken into account as
the Peierls phase in the hopping integral.
We replace
\begin{align}
	\gamma {\rm e}^{{\rm i} 2 \pi \mu \dnu_j}
	\rightarrow
	\gamma {\rm e}^{{\rm i} 2 \pi \mu \dnu_j}
	       \exp \left( {\rm i} 2 \pi \phi \frac{a_{\rm CC} \cos \Theta_j}
			{\pi d_{\rm t}} \right)
\end{align}
in $H_\mu$ in Eq.\ \eqref{eq:H0} and $H_{\mu} (k)$ in Eq.\ \eqref{eq:H0_k}.
Here,
\begin{align}
	\phi = \frac{B \, \pi (d_{\rm t}/2)^2}{h/e}
		\label{eq:phi}
\end{align}
is the AB phase, or number of flux quanta penetrating the tube,
$a_{\rm CC} = a / \sqrt3$ is the bond length $|\bm{\Delta}_j|$,
and $\Theta_j$ is the angle between $\bm{\Delta}_j$
and $\bm{C}_{\rm h}$ on the graphene sheet:
$\Theta_j = \theta - (5 \pi / 6) + (2 \pi / 3) j$.

As a result, $f_\mu (k)$ in Eq.\ \eqref{eq:f0} changes to
\begin{align}
	f_\mu (k; \phi) &= \sum_{j = 1}^3 
	{\rm e}^{{\rm i} 2 \pi \mu \dnu_j / d}
		\exp \left( {\rm i} 2 \pi \phi
			\frac{a_{\rm CC} \cos \Theta_j} {\pi d_{\rm t}} \right)
	{\rm e}^{{\rm i} k a_z \dell_j}
	\label{eq:f0_phi}
\end{align}
in a magnetic field.
$f_\mu (k; \phi)$ can be zero even for semiconducting SWNTs,
that is, the band gap is closed
at $|\phi| = \phi^*=1/3$ \cite{Ajiki1993}.
When $|\phi| \ne \phi^*$, $w_\mu$ in Eq.\ \eqref{eq:w} can be
defined in terms of $f_\mu (k; \phi)$.
As we will show later, a sudden change in $w_\mu$
takes place at $|\phi| = \phi^*=1/3$, which corresponds to the
topological phase transition.

Note that only the decimal part of $\phi$ is
physically significant.
$\phi \rightarrow \phi + 1$
compensates with $\mu \rightarrow \mu - 1$ in the
definition of angular momentum.
Therefore, we can restrict ourselves to $0 \leq \phi < 1$ or
$-1/2 \leq \phi < 1/2$, depending on the situations.

\section{Topological classification of semiconducting nanotube
	\label{sec:analytical}}

Now we topologically classify semiconducting SWNTs.
The winding number $w_\mu$
is analytically evaluated as a function of
chirality $(n, m)$
and magnetic field $B$ in the axial direction.
The winding number $w_\mu$ in Eq.\ \eqref{eq:w} can be
interpreted as the number of times that
$f_\mu (k)$ or $f_\mu (k; \phi)$ circulates around
the origin on the complex plane when $k$ runs through the 1D BZ,
$
	-\pi \leq k a_z < \pi
$.

\begin{figure}[t] \begin{center}
\includegraphics[width=8cm]{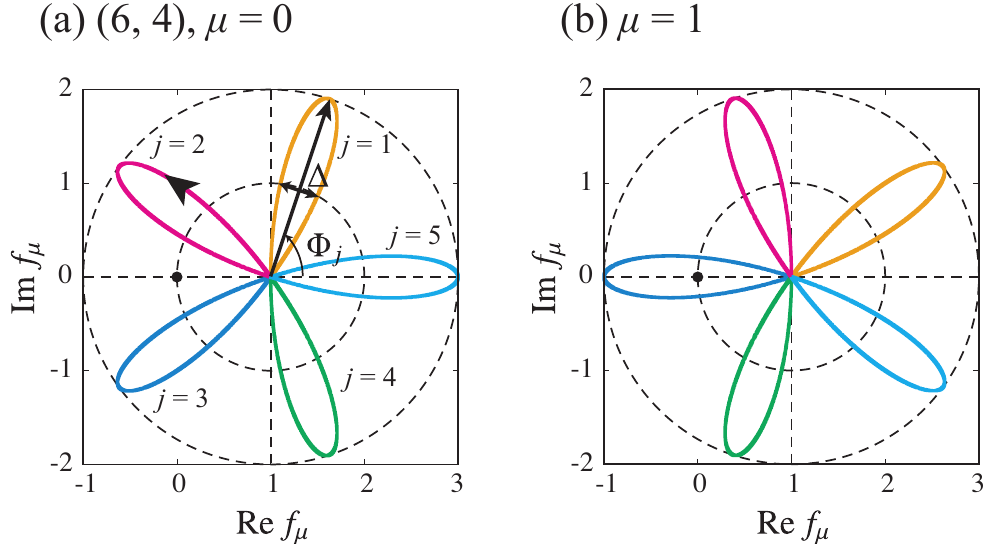}
\caption{
	(Color online)
	$f_\mu (k)$ on the complex plane for
	the (6, 4)-SWNT with (a) $\mu = 0$ and (b) $\mu = 1$.
	$d = 2$ and we choose $p = q = -1$.
	It draws a flower-shaped trajectory centered at $z = 1$,
	with petal $j=1$ to $5$.
	The distance from $z = 1$ takes its maximum, $2$, when
	the argument measured from $z = 1$ is $\Phi_j$ in
	Eq.\ \eqref{eq:Phi_j}, whereas it is 1 for $\Phi_j \pm \Delta / 2$.
}
\label{fig:flower-semicond}
\end{center} \end{figure}

\subsection{Analysis without magnetic field}

We begin with the case in the absence of a
magnetic field (AB phase $\phi = 0$).
From Eq.\ \eqref{eq:f0}, we obtain
\begin{align}
	f_\mu (k)  = 1 + 2 &\cos \left[
		\frac{n + m}{2d} k a_z - \frac{\pi \mu}{d} (p + q) \right]
		\nonumber \\
	\times &\exp \left\{ {\rm i} \left[
		\frac{n - m}{2d} k a_z + \frac{\pi \mu}{d} (- p + q) \right]
	\right\}.
		\label{eq:f_over_gamma}
\end{align}
For armchair SWNTs of $n = m$
[$d = n$ in Eq.\ \eqref{eq:def_d} and
$p = 1$ and $q = 0$
in Eq.\ \eqref{eq:pq}],
Eq.\ \eqref{eq:f_over_gamma} indicates
a line segment on the complex plane.
For SWNTs other than armchair type,
$
	f_\mu (k)
$
draws a ``flower-shaped'' closed loop,
as depicted for the $(6, 4)$-SWNT with $\mu = 0$ and $1$
in Figs.\ \ref{fig:flower-semicond}(a) and \ref{fig:flower-semicond}(b),
respectively.
We can see that the former does not circulate the origin, whereas
the latter does.
This results in $w_{\mu = 0} = 0$ and $w_{\mu = 1} = 1$,
respectively.
In general, the trajectory is centered at $z = 1$, and
$|f_\mu (k) - 1|$ takes the maximum value, $2$, when
\begin{align}
	\arg \left[ f_\mu (k) - 1 \right]
	= \frac{2 \pi d}{n + m} \left( j - \frac{\mu}{d} \right)
	\equiv \Phi_j,
		\label{eq:Phi_j}
\end{align}
with $j = 1, 2, \dots, \frac{n + m}{d}$.\footnote{
	$|f_\mu (k) - 1| = 2$ when
	$
		k a_z = \frac{2\pi d}{n + m} \left[
			j' + \frac{\mu}{d} (p + q)
		\right]
	$
	with $j' = 1, 2, \ldots, \frac{n + m}{d}$.
	Then,
	$
		\arg \left[ f_\mu (k) - 1 \right]
		= j' \pi + \frac{n - m}{2d} k a_z
			+ \frac{\pi \mu}{d} (-p + q)
		= \frac{2 \pi d}{n + m} \left(
			\frac{n}{d} j' - \frac{\mu}{d} \right)
	$.
	Since $\frac{n + m}{d}$ and $\frac{n}{d}$ are mutually prime,
	$j = \frac{n}{d} j'$ can take any integer between
	$1$ and $\frac{n + m}{d}$ with
	$j' = 1, 2, \ldots, \frac{n + m}{d}$.
	This justifies $\Phi_j$ in Eq.\ \eqref{eq:Phi_j}.
	In a similar manner, we can show that
	the argument is $\Phi_j \pm \Delta / 2$
	when $|f_\mu (k) - 1| = 1$.
}
Note that $0 < \Phi_j \le 2 \pi$ for $0 \leq \mu < d$.
$|f_\mu (k) - 1| = 1$ when
$\arg [ f_\mu (k) - 1 ] = \Phi_j \pm \Delta / 2$ with
$
	\Delta = \frac{2 \pi}{3} (n - m) / (n + m)
$.
Therefore, the $j$th ``petal'' surrounds the origin when
$
	\Phi_j - \Delta / 2 < \pi < \Phi_j + \Delta / 2
$,
that is,
\begin{align}
	\frac{n + 2m}{3d} + \frac{\mu}{d} < j
		< \frac{2n + m}{3d} + \frac{\mu}{d}.
		\label{eq:cond_j}
\end{align}
$w_\mu$ is equal to the number of integers $j$ that satisfy
Eq.\ \eqref{eq:cond_j} for given $(n, m)$ and $\mu$.
We evaluate $w_\mu$ for semiconducting SWNTs
[${\rm mod} (2n + m, 3) \ne 0$] in Table \ref{tab:w},
which are categorized according to
${\rm mod} (\frac{2n + m}{d}, 3) = 1$ or $2$.
The table also includes $w_\mu$ for metallic SWNTs with
${\rm mod} (2n + m, 3) = 0$ when the number of
times that $f_\mu (k)$ passes the origin is neglected.

\begin{table}[t]
\caption{
	The winding number $w_\mu$ determined by
	the number of times that $f_\mu (k)$ winds around the
	origin on the complex plane when $k$ runs through the 1D BZ.
	We assume $0 \leq m \leq n$, and $d = \gcd(n, m)$.
	$\mu$ is an integer (angular momentum) in the absence of
	a magnetic field while it is a real number in the presence of
	an axial magnetic field (see text in Sec.\ \ref{sec:analytical} B).
	For metallic SWNTs, we disregard the number of times that
	$f_\mu (k)$ passes the origin.
}
\label{tab:w}
\begin{tabular}{ll}
	\hline \hline
	Type $\quad$ & $w_\mu$ \\
	\hline
	\multicolumn{2}{l}{
	${\rm mod}\left( \frac{2n + m}{d}, 3 \right) = 1$
	(semiconductor or metal-1)} \\
	& $ \begin{cases}
		\frac{(n - m)/d - 2}{3} &
		\left( \frac{d}{3}  \leq \mu \leq \frac{2d}{3} \right) \\
		\frac{(n - m)/d + 1}{3} &
		\left( 0 \leq \mu < \frac{d}{3} ~{\rm or}~
			\frac{2d}{3} < \mu < d \right)
	\end{cases}$ \\
	\multicolumn{2}{l}{
	${\rm mod}\left( \frac{2n + m}{d}, 3 \right) = 2$
	(semiconductor or metal-1)} \\
	& $ \begin{cases}
		\frac{(n - m)/d + 2}{3} &
		\left( \frac{d}{3} < \mu < \frac{2d}{3} \right) \\
		\frac{(n - m)/d - 1}{3} &
		\left( 0 \leq \mu \leq \frac{d}{3} ~{\rm or}~
			\frac{2d}{3} \leq \mu < d \right)
	\end{cases}$ \\
	\multicolumn{2}{l}{
	${\rm mod}\left( \frac{2n + m}{d}, 3 \right) = 0$ and $n \neq m$
	(metal-2 other than armchair)} \\
	& $ \begin{cases}
		\frac{n - m}{3d} - 1 & (\mu = 0) \\
		\frac{n - m}{3d}     & (0 < \mu < d)
	\end{cases}$ \\
	\multicolumn{2}{l}{$n = m$ (metal-2 of armchair type)} \\
	& $0 \qquad (0 \leq \mu < d)$ \\
	\hline \hline
\end{tabular}
\end{table}

\subsection{Analysis with finite magnetic field}

When the axial magnetic field is present,
the trajectory of
$
	f_\mu (k; \phi)
$
is examined to evaluate $w_\mu$.
We obtain $\Phi_j$ in Eq.\ \eqref{eq:Phi_j}
with $\mu$ replaced by $\mu + \phi$,
which means that the trajectory for each $\mu$ is
rotated around $z = 1$ on the complex plane
	\cite{Zang2018}.
As we mentioned earlier in Sec.\ \ref{sec:mag},
only the decimal part of $\phi$ is physically
meaningful because
$
	\phi \rightarrow \phi' = \phi - \lfloor \phi \rfloor
$
is equivalent to
$
	\mu \rightarrow \mu' = \mu + \lfloor \phi \rfloor
$
with $\lfloor x \rfloor$ being the maximum integer not 
exceeding $x$.
Thus we can make the same analysis as in the previous subsection
with $\mu' = 0, 1, 2, \ldots, d - 1$, $0 \leq \phi' < 1$,
and $0 < \Phi_j \leq 2 \pi$. Then the replacement of
$\mu$ by $\mu' + \phi'$ yields the same result as in Table \ref{tab:w}.

\subsection{Edge states and topological order}

By summing up $w_\mu$ in Eq.\ \eqref{eq:bulk_edge} carefully,
we obtain the number of edge states $N_{\rm edge}$,
as shown in Table \ref{tab:N_edge}.
Here, we assume $-1/2 \leq \phi < 1/2$
($\mu$ should be shifted accordingly).
The semiconducting SWNTs are categorized into type-1 and
type-2 for ${\rm mod}(2n + m, 3) = 1$ or $2$.\footnote{A comment
is given for the classifications in Tables \ref{tab:w} and \ref{tab:N_edge}.
Semiconducting SWNTs belong to type-1 in Table \ref{tab:N_edge} when
${\rm mod}(\frac{2n + m}{d}, 3) = 1$ and ${\rm mod}(d, 3) = 1$,
or ${\rm mod}(\frac{2n + m}{d}, 3) = 2$ and ${\rm mod}(d, 3) = 2$
in Table \ref{tab:w}. They belong to type-2 otherwise.}
The results for $N_{\rm edge}$ indicate that
(i) the semiconducting SWNTs other than $n=m+1$
are topological nontrivial in the absence of a magnetic field
(AB phase $\phi=0$) and (ii) all the semiconducting SWNTs
show the topological phase transition at $|\phi|=\phi^*=1/3$
when the energy gap is closed \cite{Ajiki1993}.
Note that $|\phi|=1/3$ corresponds to the magnetic field
of more than $100~{\rm T}$ when the tube diameter
$d_{\rm t} \sim 1~{\rm nm}$.
Table \ref{tab:N_edge} also includes the results for metallic SWNTs,
that are topological insulators except for the armchair
nanotubes, irrespectively of the magnetic field, as discussed
in the next section.

\begin{figure}[t] \begin{center}
\includegraphics{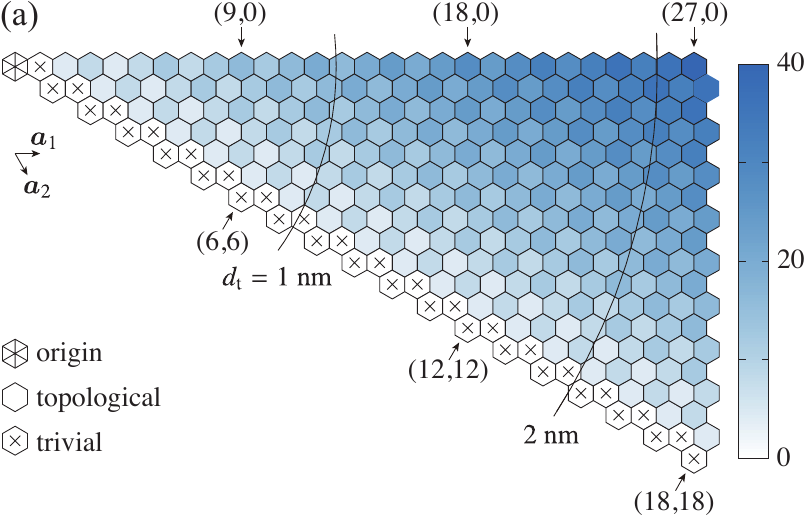}
\includegraphics{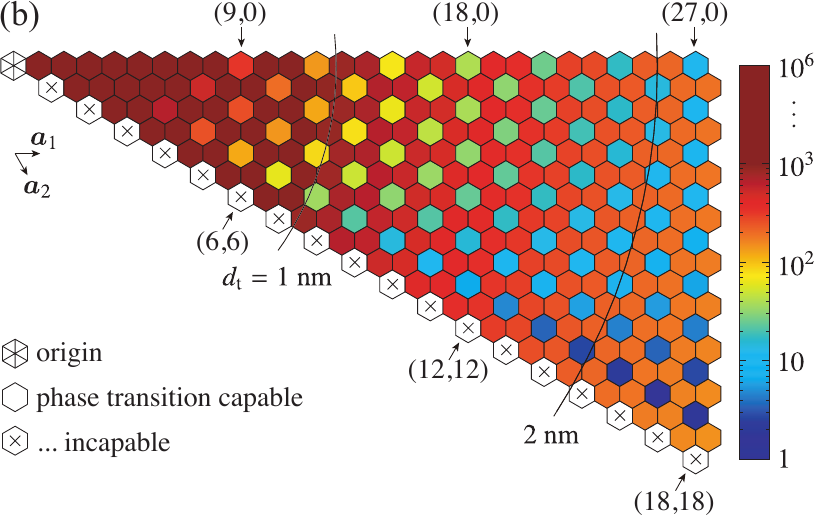}
\caption{
	(Color online)
	(a) Number of edge states in the absence of a
	magnetic field ($B = 0$) and
	(b) critical magnetic field $B^*$
	of topological phase transition at which
	the number of edge states changes discontinuously.
	A hexagon from the leftmost one indicates the
	chiral vector
	$
		\bm{C}_{\rm h} = n \bm{a}_1 + m \bm{a}_2
	$, where
	$\bm{a}_1$ and $\bm{a}_2$ are the primitive
	lattice vectors of graphene, shown in Fig.\ \ref{fig:lattice}.
}
\label{fig:class}
\end{center} \end{figure}

Figure \ref{fig:class}(a) illustrates the number of
edge states at $B = 0$, where a hexagon from the
leftmost one indicates the chiral vector
$\bm{C}_{\rm h} = n \bm{a}_1 + m \bm{a}_2$.
Almost all the SWNTs have edge states except for the
semiconducting SWNTs with $n = m + 1$ and
metallic ones of armchair type.
Figure \ref{fig:class}(b) shows the critical magnetic
field for the topological phase transition, where the number of
edge states changes discontinuously.
The critical magnetic field should be experimentally accessible
for metallic SWNTs with $d_{\rm t} \gtrsim 1~{\rm nm}$
(see Sec.\ \ref{sec:metal}).

\begin{table}[t]
\caption{
	Number of edge states $N_{\rm edge}$ in the $(n, m)$-SWNT.
	We assume $0 \leq m \leq n$.
	The number of flux quanta $\phi$ is restricted to
	$-1/2 \leq \phi < 1/2$.
	$\Theta(x) = 1$ ($0$) for $x > 0$ ($x < 0$).
}
\label{tab:N_edge}
\begin{tabular}{ll}
	\hline \hline
	Type $\quad$ & Number of edge states $N_{\rm edge}$ \\
	\hline
	\multicolumn{2}{l}{Semiconductor type-1
		[${\rm mod}(2n + m, 3) = 1$]} \\
	& $\begin{cases}
		4 \frac{n - m + 1}{3} &
		(0 \leq |\phi| < \frac13) \\
		4 \frac{n - m - 2}{3} &
		(\frac13 < |\phi| \leq \frac12) \\
	\end{cases}$ \\
	\multicolumn{2}{l}{Semiconductor type-2
		[${\rm mod}(2n + m, 3) = 2$]} \\
	& $ \begin{cases}
		4 \frac{n - m - 1}{3} &
		(0 \leq |\phi| < \frac13) \\
		4 \frac{n - m + 2}{3} &
		(\frac13 < |\phi| \leq \frac12) \\
	\end{cases}$ \\
	\multicolumn{2}{l}{Metal other than armchair
		[${\rm mod}(2n + m, 3) = 0$ and $n \neq m$]} \\
	& $ 4 \left( \frac{n - m}{3} - 1 \right) + 2 \sum_{\tau, s} \Theta(
	\Delta k_c + \tau s \Delta k_{\rm so} + \tau \Delta k_\phi)$ \\
	\multicolumn{2}{l}{Metal of armchair type ($n = m$)} \\
	& 0 \\
	\hline \hline
\end{tabular}
\end{table}

The number of edge states per diameter approaches
$N_{\rm edge} / d_{\rm t} \rightarrow
4 (n - m) / (3 d_{\rm t})$
as $d_{\rm t}$ increases.
This agrees with the result in Ref.\
	\cite{Akhmerov2008}
for the edge states of graphene.
We thus obtain an asymptotic form of $N_{\rm edge}$ for
large $d_{\rm t}$,
\begin{align}
	N_{\rm edge} \simeq \frac{8 \pi d_{\rm t}}{3a}
		\cos \left( \theta + \frac{\pi}{3} \right).
\end{align}
This should be useful when a nanotube of large diameter is
examined in a continuum approximation.

\section{Analysis for metallic nanotube
	\label{sec:metal}}

In this section, we discuss the topology of metallic SWNTs
with
$
	{\rm mod} (2n + m, 3) = 0.
$
Without the curvature-induced effects, a band of
angular momentum $\mu_{+}$ ($\mu_{-}$) passes the Dirac
point $K$ ($K'$) with wave number $k_{+}$ ($k_{-}$) on the
graphene sheet:
\begin{align}
	\mu_\pm &= \pm \frac{2n + m}{3}
	= \pm \frac{d}{3} {\rm mod} \left( \frac{2n + m}{d}, 3 \right)
	\quad ({\rm mod}~d),
		\label{eq:mu_pm} \\
	k_\pm &= \pm \frac{2 \pi}{3a_z} (2p + q),
		\label{eq:k_pm}
\end{align}
	\cite{Izumida2016}.
Metallic SWNTs are classified into metal-1
for ${\rm mod}(\frac{2n + m}{d}, 3) \ne 0$ and metal-2
for ${\rm mod}(\frac{2n + m}{d}, 3) = 0$.
$\mu_+ \neq \mu_-~({\rm mod}~d)$ in the former, whereas
$\mu_+ = \mu_- = 0$ in the latter.

In order to describe the narrow energy gap in metallic SWNTs,
we further extend our 1D lattice model to include
the curvature-induced effects besides the AB effect in
a magnetic field.
As seen in Appendix \ref{app:derivation},
our model is constructed so as to reproduce
the effective Hamiltonian for $\bm{k} \cdot \bm{p}$ theory, which
describes the curvature-induced effects and SO interaction
	\cite{Izumida2009},
in the vicinity of $k_\pm$ with angular momentum $\mu_\pm$.

\subsection{1D lattice model with curvature effects}

\begin{figure}[t] \begin{center}
\includegraphics{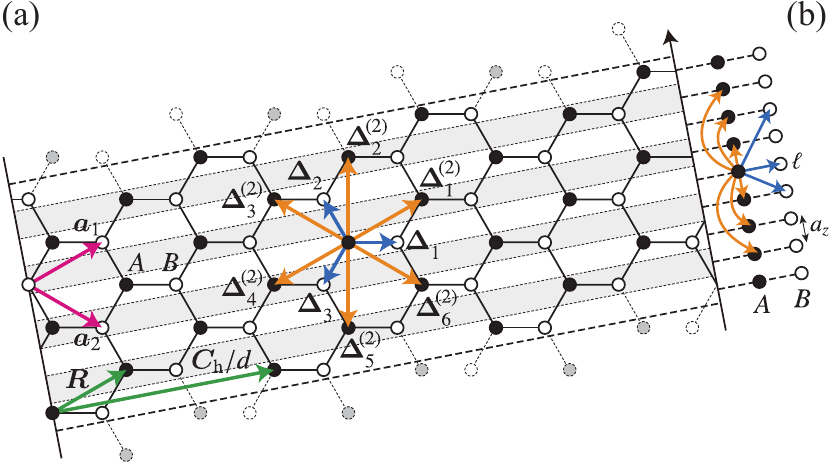}
\caption{
	(Color online) (a)
	An extension of Fig.\ \ref{fig:lattice}(a) to include the hopping to
	the second-nearest-neighbor atoms.
	The three vectors
	$\bm{\Delta}_j$
	$(j = 1, 2, 3)$
	connect the nearest-neighbor atoms,
	whereas the six vectors
	$\bm{\Delta}^{\!\!(2)}_j$ $(j = 1, 2, \ldots, 6)$ connect
	the second-nearest-neighbor ones.
	(b) An extended 1D lattice model to describe the metallic SWNTs.
}
\label{fig:lattice_2nd}
\end{center} \end{figure}

The effective Hamiltonian for $\bm{k} \cdot \bm{p}$ theory is
given by
\begin{align}
	H = &\sum_{\bm{r}_A, s} \sum_{j = 1}^3
		\left( \gamma^{(1)}_{s, j} \,
		c^{\, s \, \dagger}_{\bm{r}_A}
		c^{\, s}_{\bm{r}_A + \bm{\Delta}_j} + {\rm H.c.}
		\right)
		\nonumber \\
	+ &\sum_{\sigma=A,B} \sum_{\bm{r}_\sigma, s} \sum_{j = 1}^6
		\gamma^{(2)}_{s, j} \,
		c^{\, s \, \dagger}_{\bm{r}_\sigma}
		c^{\, s}_{\bm{r}_\sigma + \bm{\Delta}^{\!\!(2)}_j}
		\label{eq:H2d}
\end{align}
with $c^{\, s}_{\bm{r}}$ being the field operator for a $\pi$
electron with spin $s$ at atom of position $\bm{r}$
	\cite{Okuyama2017}.
The quantization axis for spin $s = \pm 1$ is chosen in the tube direction
	\cite{Izumida2009}.
This model consists of anisotropic and spin-dependent
hopping integrals to
the nearest-neighbor atoms and those to the second nearest neighbors.
As mentioned in Sec.\ \ref{sec:semicond} B,
the former connects $A$ and $B$ atoms
that are depicted by three vectors
$
	\bm{\Delta}_j
$
$(j = 1, 2, 3)$,
whereas the latter connects atoms of the same species
indicated by six vectors
$
	\bm{\Delta}^{\!\!(2)}_j
$
$(j = 1, 2, \ldots, 6)$ in Fig.\ \ref{fig:lattice_2nd}(a).
The explicit forms of hopping integrals,
$
	\gamma^{(i)}_{s,j}
$
$(i = 1, 2)$,
are provided in Appendix \ref{app:derivation}.

As described in Sec.\ \ref{sec:semicond} A, we use a
set of primitive lattice vectors, $\bm{R}$ and $\bm{C}_{\rm h} / d$.
By performing the Fourier transformation for
the $\nu$ coordinate in Eqs.\ \eqref{eq:r_A} and \eqref{eq:r_B}, we obtain
$
	H = \sum_{\mu = 0}^{d - 1} \sum_{s = \pm} H_{\mu, s}
$
with
\begin{align}
	H_{\mu, s} = &\sum_{\ell} \sum_{j = 1}^3 \left( \gamma^{(1)}_{s, j}
		{\rm e}^{{\rm i} 2 \pi \mu \dnu_j / d}
		c^{\, \mu, s \, \dagger}_{A, \ell}
		c^{\, \mu, s}_{B, \ell + \dell_j}
		+ {\rm H.c.} \right)
		\nonumber \\
	+ &\sum_{\sigma=A,B} \sum_{\ell} \sum_{j = 1}^6
		\gamma^{(2)}_{s, j}
		{\rm e}^{{\rm i} 2 \pi \mu \dnutwo_j / d}
		c^{\, \mu, s \, \dagger}_{\sigma, \ell}
		c^{\, \mu, s}_{\sigma, \ell + \delltwo_j},
		\label{eq:H1d}
\end{align}
where $c^{\, \mu, s}_{\sigma, \ell}$ is the field operator
of an electron with angular momentum $\mu$, spin $s$, and
at sublattice $\sigma$ of
site index $\ell$ in Fig.\ \ref{fig:lattice_2nd}(b).
This is an extended 1D lattice model
(see Appendix \ref{app:derivation}
for $\delltwo_j$ and $\dnutwo_j$).

\subsection{Bulk properties}

For the bulk system,
the Fourier transformation of $H_{\mu, s}$ along the $\ell$ direction
yields the two-by-two Hamiltonian,
\begin{align}
	H_{\mu, s} (k) &= \epsilon_{\mu, s} (k; \phi)+\gamma
		\begin{bmatrix}
		0 & f_{\mu, s} (k; \phi) \\
		f_{\mu, s}^* (k; \phi) & 0
	\end{bmatrix},
		\label{eq:H_k}
\end{align}
in the sublattice space for the 1D BZ,
$
	-\pi \leq k a_z < \pi
$,
where
\begin{align}
	f_{\mu, s} (k; \phi) &= \frac{1}{\gamma}\sum_{j = 1}^3
		\gamma^{(1)}_{s, j}
		{\rm e}^{{\rm i} 2 \pi \mu \dnu_j / d}
		{\rm e}^{{\rm i} k a_z \dell_j},
		\label{eq:f} \\
	\epsilon_{\mu, s} (k; \phi) &=
		\sum_{j = 1}^6 \gamma^{(2)}_{s, j}
		{\rm e}^{{\rm i} 2 \pi \mu \dnutwo_j / d}
		{\rm e}^{{\rm i} k a_z \delltwo_j}.
		\label{eq:epsilon}
\end{align}
The dispersion relation for subband $(\mu, s)$ is given by
\begin{align}
	E_{\mu, s} (k; \phi) &=
		\epsilon_{\mu, s} (k; \phi)
		\pm |\gamma f_{\mu, s} (k; \phi)|.
		\label{eq:energy_metal}
\end{align}
The system is an insulator when
$
	|\epsilon_{\mu, s}(k; \phi)| < |\gamma f_{\mu, s}(k; \phi)|
$
in the whole BZ.
Thanks to the curvature-induced fine structure,
this condition is satisfied
even for metallic SWNTs except in the vicinity of
$|\phi| = \phi^*$,
where the band gap is closed by a magnetic field.
Then positive and negative $E_{\mu, s} (k)$'s form
the conduction and valence bands, respectively.
It should be mentioned that $\phi^* \ll 1/3$ in metallic SWNTs,
which corresponds
to the magnetic field of a few Tesla
	\cite{Okuyama2017}.

\subsection{Winding number and bulk-edge correspondence}

For any SWNT with finite band gap,
we can define the winding number as
\begin{align}
	w_{\mu, s} = \frac{1}{2 \pi} \oint_{\rm BZ}
		{\rm d} \arg f_{\mu, s} (k; \phi),
		\label{eq:w_metal}
\end{align}
for subband $(\mu, s)$.
Strictly speaking, it is a topological invariant
only if the sublattice symmetry holds
	\cite{Wen1989, Asboth2015}:
$
	\epsilon_{\mu, s} (k; \phi) = 0
$.
However, as far as the system is an insulator, i.e.,
$
	|\epsilon_{\mu, s}(k; \phi)| < |\gamma f_{\mu, s}(k; \phi)|
$
in the whole BZ, it is well defined.
We discuss the topology of metallic SWNTs using $w_{\mu, s}$
in Eq.\ \eqref{eq:w_metal} except for the vicinity of $|\phi|=\phi^*$.

The bulk-edge correspondence in Eq.\ \eqref{eq:bulk_edge} is
generalized to
\begin{align}
	N_{\rm edge} = 2 \sum_{\mu = 0}^{d - 1} \sum_{s = \pm}
		|w_{\mu, s}|
		\label{eq:bulk_edge_metal}
\end{align}
in terms of $w_{\mu, s}$.
The proof of this relation is given in Appendix \ref{app:EOM}.
Although the energy levels of edge states are slightly deviated from
$E_{\rm F} = 0$ in the presence of $\epsilon_{\mu, s} (k; \phi)$,
they are within the band gap as long as the gap remains finite.

\subsection{Classification with curvature effects}

\begin{figure}[t] \begin{center}
\includegraphics{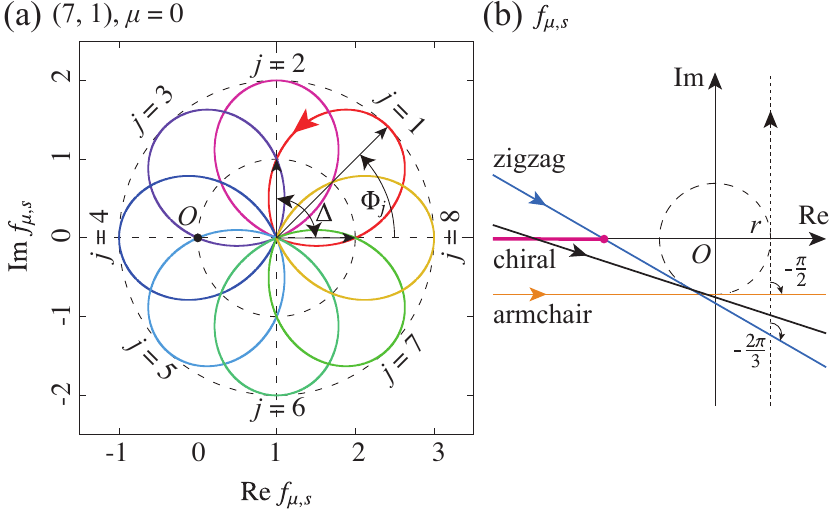}
\caption{
	(Color online)
	(a) $f_{\mu,s} (k)$ on the complex plane for
	$\mu = 0$ in a metallic SWNT with chirality $(7, 1)$.
	$d = 1$, $p = 1$, and $q = 0$. Petals $j=3$ and $5$ pass the
	origin in the absence of curvature effects, whereas 
	petal $j=4$ rounds the origin.
	(b) $f_{\mu, s} (k)$ around the origin on the complex plane
	for $\mu =\mu_{+}$ which passes the $K$ point, in metallic SWNTs
	with chiral angle $\theta$.
	$
		r = (\Delta k_c + s \Delta k_{\rm so} + \Delta k_\phi)
			(\sqrt3 a / 2) > 0
	$.
	The dotted line intersects the real axis orthogonally at $r$.
	The trajectory for a zigzag SWNT ($\theta = 0$) is obtained by
	rotating it by $- 2 \pi / 3$
	around the origin
	(blue line),
	whereas that for an armchair SWNT ($\theta = \pi/6$) is obtained by
	rotating it by $- \pi / 2$
	(orange line).
	Therefore the intercept of the real axis is always negative
	(pink segment) for SWNTs of $0 \leq \theta < \pi/6$.
}
\label{fig:flower}
\end{center} \end{figure}

Now we come to
classify the metallic SWNTs.
$f_\mu (k; \phi)$ defined in
Secs.\ \ref{sec:semicond} and \ref{sec:mag}
passes the origin on the complex plane
for $\mu = \mu_\pm$ (at $k=k_\pm$) corresponding to the
Dirac points in the absence of curvature-induced effects.
We evaluate $w_{\mu, s}$ using Eq.\ \eqref{eq:w_metal}
around the origin while we can use the results in Table \ref{tab:w}
otherwise since the topological nature does not change by a
small perturbation.
As an example, we show $f_\mu (k; \phi)$ on the complex
plane for $\mu=0$ in the (7,1)-SWNT (metal-2 with
$\mu_+=\mu_-=0$) in Fig.\ \ref{fig:flower}(a).
Petals $j=3$ and $5$ go through the origin in the absence of
curvature effects, whereas  petal $j=4$ winds the origin.
The latter yields $w_\mu = 1$ in Table \ref{tab:w}.
The contribution from the
former is discussed in the following.

In the vicinity of origin on the complex plane,
\begin{align}
	&f_{\mu_{\pm}, s} (k; \phi) \simeq \frac{\sqrt3}{2} a \,
		{\rm e}^{\pm {\rm i} (\theta - \frac{2\pi}{3})}
		\bigl[ (\Delta k_c \pm \Delta k_{\rm so}
			\pm \Delta k_\phi)
		\nonumber \\
	& \quad + {\rm i} (k - k_\tau \mp \Delta k_z) \bigr].
		\label{eq:f_metal}
\end{align}
from Eq.\ \eqref{eq:f_near_Dirac} in Appendix \ref{app:derivation}.
Here, $\Delta k_\phi$ represents the AB effect in a magnetic field,
$\Delta k_c$ and $\Delta k_z$ stem from the mixing between $\pi$ and
$\sigma$ orbitals, and $\Delta k_{\rm so}$ is due to the SO interaction
(see Appendix \ref{app:derivation}). Equation \eqref{eq:f_metal}
indicates a straight line made by the rotation of angle
$\pm (\theta - \frac{2\pi}{3})$ around the origin,
from a straight line which
intersects orthogonally the real axis at
$
	r = (\Delta k_c \pm s \Delta k_{\rm so} \pm \Delta k_\phi)
		(\sqrt3 a/2)
$.
It gives rise to the winding number
when the line intercepts the real axis in the negative part.
For armchair SWNTs of $\theta = \pi/6$,
this condition is never satisfied since
the line is parallel to the real axis.
For the other metallic SWNTs,
the condition holds if $r > 0$,
as shown in Fig.\ \ref{fig:flower}(b).

In consequence,
we obtain the complete expression for $w_{\mu, s}$ for
metallic SWNTs.
For $\mu \neq \mu_\pm$,
\begin{align}
	w_{\mu, s} = w_\mu
	\label{eq:w_mu_s0}
\end{align}
in Table \ref{tab:w}.
For $\mu = \mu_\pm$,
\begin{align}
	w_{\mu_\pm, s} = w_{\mu_\pm}
	+ \Theta(\Delta k_c \pm s \Delta k_{\rm so} \pm \Delta k_\phi),
		\label{eq:w_mu_s}
\end{align}
where $w_{\mu_\pm}$ is given by Table \ref{tab:w} and
$\Theta(x) = 1$ ($0$) for $x > 0$ ($x < 0$).
This explains the topological phase transition
at $|\phi| \ll 1/3$,
which was demonstrated in Ref.\ \cite{Okuyama2017},
for the following reason.
$\Delta k_\phi$ is proportional to $B$ along the tube axis,
$\Delta k_\phi = - eBd_{\rm t}/(4 \hbar)$,
in Eq.\ \eqref{eq:k_phi} in Appendix \ref{app:derivation}.
For metallic SWNTs other than the armchair,
$
	\Delta k_c \pm s \Delta k_{\rm so} \simeq \Delta k_c > 0
$
and thus Eq.\ \eqref{eq:w_mu_s} yields $w_{\mu_\pm, s} =
w_{\mu_\pm}+1$ at $B=0$.
When $B$ is increased beyond $B^*$, which satisfies
$
	\Delta k_c + s \Delta k_{\rm so} + \Delta k_\phi = 0
$, $w_{\mu_+, s}$ becomes $w_{\mu_+}$.

We obtain the number of edge states $N_{\rm edge}$
through Eq.\ \eqref{eq:bulk_edge_metal}
by the summation of $w_{\mu_\pm, s}$ in Eqs.\ \eqref{eq:w_mu_s0}
and \eqref{eq:w_mu_s}. The expression for $N_{\rm edge}$
is common for metal-1 and -2, as shown in Table \ref{tab:N_edge}.
All the metallic SWNTs but the armchair are a topological insulator
in the absence of a magnetic field [Fig.\ \ref{fig:class}(a)] and show the
topological phase transition at $B=B^*$ [Fig.\ \ref{fig:class}(b)].
The armchair SWNTs are always topologically trivial:
They are forbidden to have finite winding
numbers regardless of the strength of the magnetic field,
which is attributable to
the mirror symmetry with respect to a
plane including the tube axis
	\cite{Izumida2016}.

\section{Discussion
	\label{sec:discussion}}

We comment on the previous studies which predicted
an increase in the number of edge states in metallic SWNTs as
the magnetic field increases
	\cite{Sasaki2005, Sasaki2008, Marganska2011}.
At the first sight, this seems contradictory against our results.
However, this is because they use parameters corresponding to
$\Delta k_c < 0$ in our model.
We obtain positive $\Delta k_c$ by fitting the dispersion relation with
that from the {\it ab initio} calculation known as
the extended tight-binding model
	\cite{Izumida2009}.
However, its sign is quite sensitive to
the details in the model,
and therefore it should be experimentally confirmed which sign is favorable.
Also, others theoretically predicted no topological phase transition for
metallic SWNTs
	\cite{Lin2016}.
This is due to the oversimplification with
$\Delta k_c = \Delta k_{\rm so} = 0$.

A comment should be made on the boundary condition,
which is important for the edge states in 1D topological insulators.
Our calculations have been performed for finite systems
in which a SWNT is cut by a broken line in Fig.\ \ref{fig:lattice}(a).
The angular momentum $\mu$ is
a good quantum number in this case.
This is a minimal boundary edge,
where every atom at the ends has just one dangling bond
	\cite{Akhmerov2008}.
The bulk-edge correspondence in
Eqs.\ (\ref{eq:bulk_edge}) and (\ref{eq:bulk_edge_metal})
holds only for such edges.
Some other boundary conditions result in different numbers of edge states,
as discussed in Ref.\ \cite{Izumida2016}.
Then the winding number $w_{\mu}$ is shifted from that in the case of
minimal boundary. Since the shift of $w_{\mu}$ is independent of
magnetic field \cite{Izumida2016}, the topological phase transition and
the critical magnetic field should not be influenced by the boundary
conditions. The number of edge states is changed at the transition.
For armchair SWNTs, the topological phase transition does not take
place with any boundary condition, whereas the number of edge states
may be finite.
Although the examined boundaries are limited,
we speculate that the topological phase transition is determined by
the topological nature of the bulk irrespectively of the boundaries
in general.

\section{Conclusions
	\label{sec:conclusions}}

We have classified the topology for all possible chiralities $(n,m)$
of SWNTs in the absence and presence of a magnetic field along the tube axis.
First, we have studied semiconducting SWNTs using a 1D lattice model
in Eq.\ \eqref{eq:H0} and depicted in Fig.\ \ref{fig:lattice}(b).
We have found that
(i) the semiconducting SWNTs other than $n=m+1$
are topological nontrivial in the absence of a magnetic field
and (ii) all the semiconducting SWNTs show the topological phase
transition at AB phase $|\phi|=\phi^*=1/3$. The phase transition,
however, should be hard to observe since a magnetic field of more
than $100~{\rm T}$ is required
when the tube diameter $d_{\rm t} \sim 1~{\rm nm}$.

Next, we have examined metallic SWNTs with a small band gap
using an extended 1D lattice model in Eq.\ \eqref{eq:H1d} and depicted in
Fig.\ \ref{fig:lattice_2nd}(b). Although the winding number $w_{\mu,s}$ is not
a topological invariant in the presence of $\gamma^{(2)}_{s,j}$
in Eq.\ \eqref{eq:H1d}, it is well defined except for the vicinity of
topological phase transition. Indeed we have proved the bulk-edge
correspondence for $w_{\mu,s}$ in Eq.\ \eqref{eq:bulk_edge_metal}.
We have observed that
(i) all the metallic SWNTs but the armchair type ($n=m$) are a
topological insulator in the absence of a magnetic field and show the
topological phase transition at a critical magnetic field $B^*$.
Since $B^*$ can be a few Tesla
	\cite{Okuyama2017},
the topological phase
transition could be observed for metallic SWNTs.
(ii) The armchair SWNTs are always topologically trivial.

In conclusion, the majority of SWNTs are a topological insulator
in the absence of a magnetic field and show a topological
phase transition by applying a magnetic field along the tube.
Only metallic SWNTs of armchair type are topologically trivial
regardless of the magnetic field.

\acknowledgements

The authors acknowledge fruitful discussion with
K.\ Sasaki, A.\ Yamakage, M.\ Grifoni, and R.\ Saito.
This work was partially supported by JSPS KAKENHI Grants No.\ 26220711,
No.\ 15K05118, No.\ 15H05870, No.\ 15KK0147, No.\ 16H01046, and No.\ 18H04282.

\appendix

\section{Effective lattice model for metallic SWNTs
	\label{app:derivation}}

We construct an effective 1D lattice model for
a metallic SWNT,
starting from the Hamiltonian of
$\bm{k} \cdot \bm{p}$ theory
	\cite{Izumida2009},
as discussed in Ref.\
	\cite{Okuyama2017}.
In a magnetic field $B$ in the axial direction, the
Hamiltonian in the vicinity of $K$ and $K'$ points reads
\begin{align}
	\mathcal{H}_{\tau, s} (\bm{k}) &= \hbar v_{\rm F} \Bigl[
		\bigl( k_c - \tau \Delta k_c - s \Delta k_{\rm so}
		- \Delta k_\phi \bigr) \sigma_x
		\nonumber \\
	&\quad + \tau \bigl( k_z - \tau \Delta k_z \bigr)
		\sigma_y \Bigr]
	+ \tau s \epsilon_{\rm so},
		\label{eq:H_continuum}
\end{align}
with $\sigma_x$ and $\sigma_y$ being the Pauli
matrices in the sublattice space of $A$ and $B$ species.
$s = \pm 1$ is the spin in the axial direction, whereas
$\tau = \pm 1$ represents $K$ or $K'$ valleys.
$k_c$ and $k_z$ are the circumference and axial components of
wave number measured from $K$ or $K'$ points, respectively.
$k_c$ is discretized in units of
$2\pi/|\bm{C}_{\rm h}|$ while $k_z$ is continuous.

In $\mathcal{H}_{\tau, s} (\bm{k})$,
the hybridization between $\pi$ and $\sigma$ orbitals appears as
the shift of Dirac points from $K$ or $K'$ points,
\begin{align}
	\Delta k_c = \beta' \frac{\cos 3\theta}{d_{\rm t}^2}, \quad
	\Delta k_z = \zeta  \frac{\sin 3\theta}{d_{\rm t}^2},
\end{align}
with $\beta' = 0.0436~{\rm nm}$ and $\zeta = -0.185~{\rm nm}$.
$\Delta k_c$ opens a small gap $E_{\rm g}$
except in armchair tubes ($\theta=\pi/6$).
The curvature-enhanced SO interaction yields
\begin{align}
	\Delta k_{\rm so} = \alpha'_1 V_{\rm so} \frac{1}{d_{\rm t}}, \quad
	\epsilon_{\rm so} = \alpha_2 V_{\rm so}
		\frac{\cos 3\theta}{d_{\rm t}},
\end{align}
with $\alpha'_1 = 8.8 \times 10^{-5}~{\rm meV^{-1}}$,
$\alpha_2 = - 0.045~{\rm nm}$, and $V_{\rm so} = 6~{\rm meV}$ being
the SO interaction for 2p orbitals in carbon atoms.
$\Delta k_{\rm so}$ opens the gap in armchair tubes and
gives a correction to $E_{\rm g}$ in the others.
The AB phase by the magnetic field $B$ appears as
\begin{align}
	\Delta k_\phi = - \frac{eB}{4 \hbar} d_{\rm t}.
		\label{eq:k_phi}
\end{align}
The band gap is closed at $B^*$ when
$\tau \Delta k_c + s \Delta k_{\rm so} + \Delta k_\phi=0$.
The last term in $\mathcal{H}_{\tau, s} (\bm{k})$ yields the
energy shift from $E_{\rm F}=0$,
which is assumed to be small compared with the band
gap except in the vicinity of $B=B^*$.

The 2D lattice model in Eq.\ \eqref{eq:H2d} is constructed to
reproduce $\mathcal{H}_{\tau, s} (\bm{k})$ around
the Dirac points
	\cite{Okuyama2017}.
The hopping integral $\gamma^{(1)}_{s, j}$ is given by
\begin{align}
	\gamma^{(1)}_{s, j} &= \gamma \exp \left( - {\rm i} \Delta
		k_\phi a_\text{\tiny CC} \cos \Theta_j \right) \biggl[ 1
		+ \Delta k_c a_\text{\tiny CC} \sin \Theta_j
		\nonumber \\
	&\quad - (\Delta k_z + {\rm i} s \Delta k_{\rm so})
		a_\text{\tiny CC} \cos \Theta_j \biggr],
		\label{eq:gamma1}
\end{align}
whereas $\gamma^{(2)}_{s, j}$
stems from the SO interaction as
\begin{align}
	\gamma^{(2)}_{s, j} = {\rm i} \frac{(-1)^{j+1}}{3\sqrt{3}}
		s \epsilon_{\rm so}.
		\label{eq:gamma2}
\end{align}

In a similar way to Sec.\ \ref{sec:semicond} A,
we derive the 1D lattice model in
Eq.\ \eqref{eq:H1d} from Eq.\ \eqref{eq:H2d}.
The hopping distance $\dell_j$
and phase factor $\dnu_j$ for the nearest-neighbor atoms
are given by Eq.\ \eqref{eq:Delta}.
For the second-nearest-neighbor atoms,
$\delltwo_j$ and $\dnutwo_j$ are determined from
\begin{align}
	\bm{\Delta}^{\!\!(2)}_j
	= \delltwo_j \bm{R}
		+ \dnutwo_j (\bm{C}_{\rm h} / d).
\end{align}
These quantities are provided in Table \ref{tab:Delta}.

\begin{table}[t]
\caption{
	Hopping distance and phase factor in the 1D lattice model.
	Integers $d$, $p$, and $q$ are given by Eqs.\
	\eqref{eq:def_d} and \eqref{eq:pq}.
}
\label{tab:Delta}
\begin{tabular}{cccccccccccccc}
	\cline{1-5} \cline{7-14} \noalign{\vspace{2pt}}
	\cline{1-5} \cline{7-14}
	$j$ & & ~~1~~ & ~~2~~ & ~~3~~ & ~~~~ &
	$j$ & & ~~1~~ & ~~2~~ & ~~3~~ & ~~4~~ & ~~5~~ & ~~6~~\\
	\cline{1-5} \cline{7-14}
	$\dell_j$ & & 0 & $\frac{n}{d}$ & $-\frac{m}{d}$ & &
	$\delltwo_j$ & & $\frac{m}{d}$  & $\frac{n+m}{d}$  & $\frac{n}{d}$ &
		$-\frac{m}{d}$ & $-\frac{n+m}{d}$ & $-\frac{n}{d}$ \\
	$\dnu_j$ & & 0 & $-p$ & $q$ & &
	$\dnutwo_j$ & & $-q$ & $-(p+q)$ & $-p$ & $q$ & $p + q$ & $p$ \\
	\cline{1-5} \cline{7-14} \noalign{\vspace{2pt}}
	\cline{1-5} \cline{7-14}
\end{tabular}
\end{table}

We examine low-lying states near to the
$K$ and $K'$ points.
By expanding the bulk Hamiltonian in
Eqs.\ \eqref{eq:H_k}--\eqref{eq:epsilon}
around $\mu = \mu_\tau$ and $k = k_\tau$ in
Eqs.\ \eqref{eq:mu_pm} and \eqref{eq:k_pm},
we obtain
\begin{align}
	\gamma f_{\mu, s} (k) &= \chi_\tau \hbar v_{\rm F} \bigl[ (
		k_c - \tau \Delta k_c - s \Delta k_{\rm so} - \Delta k_\phi)
	- {\rm i} \tau (k_z - \tau \Delta k_z) \bigr],
		\label{eq:f_near_Dirac} \\
	\epsilon_{\mu, s} (k) &= \tau s \epsilon_{\rm so},
\end{align}
with
$
	\chi_\tau = \tau {\rm e}^{{\rm i} \tau (\theta - 2 \pi / 3)}
$,
$
	k_c = (\mu - \mu_\tau) (2\pi / |\bm{C}_{\rm h}|)
$,
and $k_z = k - k_\tau$.\footnote{
	This agrees with the effective Hamiltonian of
	$\bm{k} \cdot \bm{p}$ theory
	in Eq.\ \eqref{eq:H_continuum},
	up to an irrelevant phase factor $\chi_\tau$.
}

The local band gap around $K$ and $K'$ points is evaluated by using
Eq.\ \eqref{eq:f_near_Dirac}.
For metallic SWNTs, $k_c$ can be zero, and therefore
band gap is determined by the curvature effects as
\begin{align}
	E_{\rm g} = \hbar v_F \left| \Delta k_c + \tau s \Delta k_{\rm so}
		+ \tau \Delta k_\phi \right|.
		\label{eq:metallic_gap}
\end{align}

\section{Proof of bulk-edge correspondence
	\label{app:EOM}}

In this Appendix, we evaluate the number of edge states from
the Schr\"odinger equation, in a similar manner to Ref.\
\cite{Izumida2016}.
For semiconducting SWNTs, edge states have energy $E = E_{\rm F} = 0$.
For metallic SWNTs, they still have $E = 0$ if we set
$\epsilon_{\mu, s} (k) = 0$ in Eq.\ \eqref{eq:epsilon}:
The topological nature is determined by
$\gamma^{(1)}_{s,j}$, whereas the energy of the
edge states is shifted by
$\gamma^{(2)}_{s,j}$ in Hamiltonian \eqref{eq:H2d}.
Here, we examine the edge states at $E = 0$, neglecting
$\gamma^{(2)}_{s,j}$.

We consider a long but finite SWNT using the 1D lattice model
with $\ell=0,1,2,\ldots,L$ in Fig.\ \ref{fig:lattice_2nd}(b).
Using the Hamiltonian in Eq.\ \eqref{eq:H1d}, the
time-independent Schr\"odinger equation is written as
\begin{align}
	H_{\mu, s} |\phi\rangle = E |\phi\rangle,
	\quad
	|\phi\rangle = \sum_{\sigma=A,B} \sum_{\ell=0}^L
	\phi_{\sigma, \ell} |\sigma, \ell; \mu, s \rangle,
		\label{eq:Schroedinger}
\end{align}
where $|\sigma, \ell; \mu, s \rangle$ is the
electronic states at sublattice $\sigma$ of
site index $\ell$
with angular momentum $\mu$ and spin $s$.
For $E = 0$, equations for $\{ \phi_{A, \ell} \}$ and
$\{ \phi_{B, \ell} \}$ are decoupled from each other.
We find that the edge states consist of $B$ sublattice ($A$ sublattice) only
around $\ell=0$ ($\ell=L$).\footnote{
	The edge states consisting of $A$
	($B$) sublattice around $\ell=0$ ($\ell=L$) do not exist because
	the number of boundary conditions $n/d$ is larger than
	the number of solutions of Eq.\ \eqref{eq:B_modes}.
}
Here, we examine the former with the boundary conditions
of $\phi_{B,-1} = \phi_{B,-2} = \ldots = \phi_{B, -m/d} = 0$
because of the hopping integrals from $\ell$ to $\ell+\dell_3=
\ell-m/d$ ($\ell \ge 0$).
The number of edge states is given by the number of roots
in the equation that decay into the tube body,
subtracted by the number of boundary conditions, $m/d$.

If we assume a wave function in a form of
$
	\phi_{\sigma, \ell} =
		\lambda^{\ell} \phi_{\sigma, 0}
$
with a complex number $\lambda$,
the equation for $B$ sublattice reads,
\begin{align}
	&\sum_{j = 1}^3 \gamma^{(1)}_{s,j}
		{\rm e}^{{\rm i} 2 \pi \mu \dnu_j / d}
		\lambda^{\dell_j} = 0.
		\label{eq:B_modes}
\end{align}

\begin{figure}[!t] \begin{center}
\includegraphics{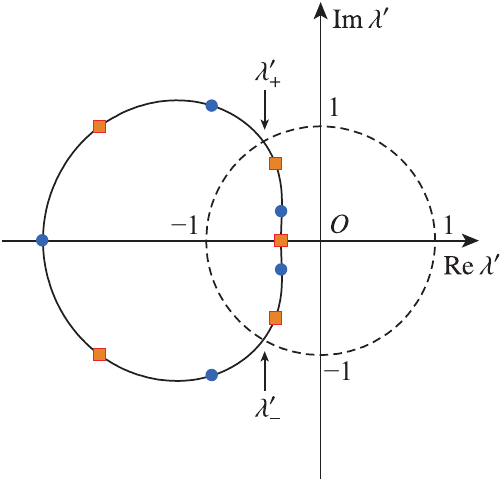}
\caption{
	(Color online)
	Zero-energy modes at $B$ sublattice in
	the $(6, 4)$-SWNT on the complex plane of
	$\lambda'$ in Eq.\ \eqref{eq:lambda_dash}.
	Circles and squares correspond to modes of
	$\mu = 0$ and $1$, respectively.
	The solid line shows Eq.\ \eqref{eq:EOM_amp},
	whereas the broken line shows a unit circle
	centered at the origin.
	They are crossed at $\lambda'_\pm = {\rm e}^{{\rm i} 2 \pi/3}$.
}
\label{fig:modes}
\end{center} \end{figure}

First, we neglect the curvature-induced effects and only
the AB phase is taken into account, i.e.,
$
	\gamma^{(1)}_{s, j} =
	\gamma \exp \left(
	{\rm i} 2 \pi \phi \frac{a_{\rm CC} \cos \Theta_j}{\pi d_{\rm t}} \right)
	= \gamma
	\exp \left( - {\rm i} \Delta k_\phi a_{\rm CC} \cos \Theta_j \right)
$.
From Eq.\ \eqref{eq:B_modes}, we obtain
\begin{align}
	1 &+ {\rm e}^{- {\rm i} 2 \pi p \mu / d}
	    {\rm e}^{- {\rm i} (n + 2 m) \phi a^2 / (4 S)}
	    \lambda^{n / d} \nonumber \\
	  &+ {\rm e}^{{\rm i} 2 \pi q \mu / d}
	    {\rm e}^{- {\rm i} (2n + m) \phi a^2 / (4 S)}
	    \lambda^{- m / d} = 0,
\end{align}
with
$
	S = \pi (d_{\rm t} / 2)^2
$
being the cross section of a SWNT.
Then a straightforward calculation yields
\begin{align}
	{\rm e}^{{\rm i} 2 \pi (\mu + \phi) / d} \left( \lambda' \right)^{m / d}
	= \left( - 1 - \lambda' \right)^{\frac{n + m}{d}},
		\label{eq:EOM}
\end{align}
where
\begin{align}
	\lambda' = {\rm e}^{- {\rm i} 2 \pi (p + q) \mu / d}
		{\rm e}^{{\rm i} (n - m) \phi a^2 / (4S)}
		\lambda^{\frac{n + m}{d}}.
		\label{eq:lambda_dash}
\end{align}
Equation \eqref{eq:EOM} yields two equations for
absolute and phase values as
\begin{align}
	&\left| \lambda' \right|^{m / d}
		= \left| - 1 - \lambda' \right|^{\frac{n + m}{d}},
		\label{eq:EOM_amp} \\
	&\frac{n + m}{d} \arg \left( - 1 - \lambda' \right)
	- \frac{m}{d} \arg \lambda' - \frac{2 \pi (\mu + \phi)}{d}
		= 2 \pi j,
		\label{eq:EOM_phase}
\end{align}
with $j$ being an arbitrary integer.
The condition Eq.\ \eqref{eq:EOM_amp} gives a closed loop
on the complex plane of $\lambda'$,
which crosses the unit circle of $\left| \lambda' \right| = 1$ at
$
	\lambda'_{\pm} = {\rm e}^{\pm {\rm i} 2 \pi / 3},
$
as shown in Fig.\ \ref{fig:modes}.
On the loop, Eq.\ \eqref{eq:EOM_phase} indicates
$\frac{n + m}{d}$ points, which are the solution of
Eq.\ \eqref{eq:EOM}.
Among them, decaying modes from $\ell = 0$
correspond to the points of
$|\lambda|=\left| \lambda' \right| < 1$.
We obtain the number of such modes by counting
integers between $j_+$ and $j_-$, where
\begin{align}
	j_+ = \frac{2n + m}{3d} - \frac{\mu + \phi}{d}, \quad
	j_- = \frac{n - m}{3d} - \frac{\mu + \phi}{d}
	\label{eq:j_pm}
\end{align}
satisfy Eq.\ \eqref{eq:EOM_phase} with $\lambda'_\pm$.

For metallic SWNTs, $j_\pm$ in Eq.\ \eqref{eq:j_pm} can be
integers for $\phi = 0$, which correspond
to $K$ and $K'$ points.
We neglect their contribution for now, and examine later.

The above-mentioned analysis yields the number of roots
of Eq.\ \eqref{eq:B_modes}.
The subtraction of $m / d$ determines the number of edge
states around $\ell = 0$ with angular momentum $\mu$,
$N_{{\rm edge},\mu}$, for each spin.
$N_{{\rm edge},\mu}$ coincides with $|w_{\mu}|$ in Table \ref{tab:w}.
The same calculation can be applied for the edge states around
$\ell = L$, which consist of $A$ sublattice. In consequence,
$4N_{{\rm edge},\,\mu}$ gives the total number of edge states,
which yields the bulk-edge correspondence in Eq.\ \eqref{eq:bulk_edge}.

As an example, Fig.\ \ref{fig:modes} depicts the edge modes around
$\ell=0$ in the $(6, 4)$-SWNT $(d = 2)$ on the $\lambda'$ plane.
Circles and squares are modes of
$\mu = 0$ and $1$, respectively.
Solid and broken lines show Eq.\ \eqref{eq:EOM_amp} and
the unit circle $|\lambda'| = 1$, respectively.
For $\mu = 0$, the number of the modes inside the broken line is two.
Since the number of boundary conditions is $m / d = 2$,
we have $2 - 2 = 0$ edge states, 
which corresponds to $w_{\mu = 0} = 0$.
For $\mu = 1$, on the other hand, we have three decaying modes
and hence one edge state, which is consistent with
$w_{\mu = 1} = 1$.

Finally, we examine the contribution from $K$ and $K'$ points
in metallic SWNTs.
Note that if we write $\lambda = {\rm e}^{({\rm i} k - \kappa) a_z}$,
Eq.\ \eqref{eq:B_modes} results in the condition of
$
	\left. f_{\mu, s} (k) \right|_{k \rightarrow k + {\rm i} \kappa} = 0
$.
Without the curvature-induced effects, a plane wave gives
its solution.
Thus we examine how the wave function is modified
when the curvature-induced effects are included.
From Eq.\ \eqref{eq:f_near_Dirac},
we find
a solution near $K$ and $K'$ points
with $\lambda = {\rm e}^{({\rm i} k - \kappa) a_z}$,
\begin{align}
	k = k_\tau + \tau \Delta k_z, \quad
	\kappa = \Delta k_c + \tau s \Delta k_{\rm so} + \tau \Delta k_\phi.
\end{align}
Then we find that there is one additional edge state
at $B$ sublattice around $\ell = 0$ if $\kappa > 0$.
This ends up the conclusion that $N_{\rm edge}$ in Table
\ref{tab:N_edge} is valid also for metallic SWNTs.
Thus the bulk-edge correspondence is established for both
semiconducting and metallic SWNTs in an arbitrary magnetic field.


\begin{thebibliography}{10}
\newcommand{\enquote}[1]{``#1''}
\providecommand{\url}[1]{\texttt{#1}}
\providecommand{\urlprefix}{URL }
\providecommand{\eprint}[2][]{e-print #2}

\bibitem{Saito1998}
R.~Saito, G.~Dresselhaus, and M.~S.~Dresselhaus, {\it {Physical properties of
  carbon nanotubes}} (Imperial College Press, London, 1998).

\bibitem{Hamada1992}
N.~Hamada, S.~Sawada, and A.~Oshiyama, \enquote{{New one-dimensional
  conductors: Graphitic microtubules}}, Phys.~Rev.~Lett.~\textbf{68},
  1579 (1992).

\bibitem{Saito1992}
R.~Saito, M.~Fujita, G.~Dresselhaus, and M.~S.~Dresselhaus,
  \enquote{{Electronic structure of graphene tubules based on C60}},
  Phys.~Rev.~B \textbf{46}, 1804 (1992).

\bibitem{Kane1997}
C.~L.~Kane and E.~J.~Mele, \enquote{{Size, Shape, and Low Energy Electronic
  Structure of Carbon Nanotubes}}, Phys.~Rev.~Lett.~\textbf{78}, 1932
  (1997).

\bibitem{Ando2000}
T.~Ando, \enquote{{Spin-orbit interaction in carbon nanotubes}},
  J.~Phys.~Soc.~Jpn \textbf{69}, 1757 (2000).

\bibitem{Klinovaja2012}
J.~Klinovaja, S.~Gangadharaiah, and D.~Loss,
  \enquote{{Electric-Field-Induced Majorana Fermions in Armchair Carbon
  Nanotubes}},
  Phys.~Rev.~Lett.~\textbf{108}, 196804 (2012).

\bibitem{Egger2012}
R.~Egger and K.~Flensberg,
  \enquote{{Emerging Dirac and Majorana fermions for carbon nanotubes with
  proximity-induced pairing and spiral magnetic field}},
  Phys.~Rev.~B \textbf{85}, 235462 (2012).

\bibitem{Sau2013}
J.~D.~Sau and S.~Tewari,
  \enquote{{Topological superconducting state and Majorana fermions in
  carbon nanotubes}},
  Phys.~Rev.~B \textbf{88}, 054503 (2013).

\bibitem{Hsu2015}
C.-H.~Hsu, P.~Stano, J.~Klinovaja, and D.~Loss,
  \enquote{{Antiferromagnetic nuclear spin helix and topological
  superconductivity in $^{13}$C nanotubes}},
  Phys.~Rev.~B \textbf{92}, 235435 (2015).

\bibitem{Izumida2016}
W.~Izumida, R.~Okuyama, A.~Yamakage, and R.~Saito, \enquote{{Angular momentum
  and topology in semiconducting single-wall carbon nanotubes}},
  Phys.~Rev.~B \textbf{93}, 195442 (2016).

\bibitem{Lin2016}
S.~Lin, G.~Zhang, C.~Li, and Z.~Song, \enquote{{Magnetic-flux-driven
  topological quantum phase transition and manipulation of perfect edge states
  in graphene tube}}, Sci.~Rep.~\textbf{6}, 31953 (2016).

\bibitem{Okuyama2017}
R.~Okuyama, W.~Izumida, and M.~Eto, \enquote{{Topological Phase Transition in
  Metallic Single-Wall Carbon Nanotube}}, J.~Phys.~Soc.~Jpn
  \textbf{86}, 013702 (2017).

\bibitem{Izumida2017}
W.~Izumida, L.~Milz, M.~Marganska, and M.~Grifoni, \enquote{{Topology and zero
  energy edge states in carbon nanotubes with superconducting pairing}},
  Phys.~Rev.~B \textbf{96}, 125414 (2017).

\bibitem{Marganska2018}
M.~Marganska, L.~Milz, W.~Izumida, C.~Strunk, and M.~Grifoni,
  \enquote{{Majorana quasiparticles in semiconducting carbon nanotubes}},
  Phys.~Rev.~B \textbf{97}, 075141 (2018).

\bibitem{Zang2018}
X.~Zang, N.~Singh, and U.~Schwingenschl, \enquote{{Topological characterization
  of carbon nanotubes}}, J.~Phys.: Cond.~Matt.~\textbf{30},
  335301 (2018).

\bibitem{Wen1989}
X.~G.~Wen and A.~Zee, \enquote{{Winding number, family index theorem, and
  electron hopping in a magnetic field}}, Nucl.~Phys.~B \textbf{316}, 641
  (1989).

\bibitem{Schnyder2009}
A.~P.~Schnyder, S.~Ryu, A.~Furusaki, and A.~W.~W.~Ludwig,
  \enquote{{Classification of topological insulators and superconductors}},
  Phys.~Rev.~B \textbf{78}, 195125 (2008).

\bibitem{Kobayashi2005}
Y.~Kobayashi, K.~Fukui, T.~Enoki, K.~Kusakabe, and Y.~Kaburagi,
  \enquote{{Observation of zigzag and armchair edges of graphite using scanning
  tunneling microscopy and spectroscopy}}, Phys.~Rev.~B \textbf{71},
  193406 (2005).

\bibitem{Sasaki2005}
K.~Sasaki, S.~Murakami, R.~Saito, and Y.~Kawazoe, \enquote{{Controlling edge
  states of zigzag carbon nanotubes by the Aharonov-Bohm flux}},
  Phys.~Rev.~B \textbf{71}, 195401 (2005).

\bibitem{Sasaki2008}
K.~Sasaki, M.~Suzuki, and R.~Saito, \enquote{{Aharanov-Bohm effect for the edge
  states of zigzag carbon nanotubes}}, Phys.~Rev.~B \textbf{77}, 045138
  (2008).

\bibitem{Marganska2011}
M.~Marga{\'{n}}ska, M.~del Valle, S.~H.~Jhang, C.~Strunk, and M.~Grifoni,
  \enquote{{Localization induced by magnetic fields in carbon nanotubes}},
  Phys.~Rev.~B \textbf{83}, 193407 (2011).

\bibitem{White1993}
C.~T.~White, D.~H.~Robertson, and J.~W.~Mintmire, \enquote{{Helical and
  rotational symmetries of nanoscale graphitic tubules}}, Phys.~Rev.~B
  \textbf{47}, 5485 (1993).

\bibitem{Jishi1993}
R.~A.~Jishi, M.~S.~Dresselhaus, and G.~Dresselhaus, \enquote{{Symmetry
  properties of chiral carbon nanotube}}, Phys.~Rev.~B \textbf{47}, 16671
  (1993).

\bibitem{Akhmerov2008}
A.~R.~Akhmerov and C.~W.~J.~Beenakker, \enquote{{Boundary conditions for Dirac
  fermions on a terminated honeycomb lattice}}, Phys.~Rev.~B \textbf{77},
  085423 (2008).

\bibitem{Ajiki1993}
H.~Ajiki and T.~Ando, \enquote{{Electronic states of carbon nanotubes}},
  J.~Phys.~Soc.~Jpn \textbf{62}, 1255 (1993).

\bibitem{Izumida2009}
W.~Izumida, K.~Sato, and R.~Saito, \enquote{{Spin-orbit interaction in single
  wall carbon nanotubes: Symmetry adapted tight-binding calculation and
  effective model analysis}}, J.~Phys.~Soc.~Jpn
  \textbf{78}, 074707 (2009).

\bibitem{Asboth2015}
J.~K.~Asb{\'{o}}th, L.~Oroszl{\'{a}}ny, and A.~P{\'{a}}lyi, {\it {A short
  course on topological insulators: Band-structure topology and edge states in
  one and two dimensions}} (Springer, Berlin, 2015).

\end{thebibliography}
\end{document}